\documentclass{aa}  
\usepackage{graphicx}
\usepackage{txfonts}
\begin{document}
   \title{Abell\,43: Longest Period Planetary Nebula Nucleus Variable\thanks{Based on observations made
 with the Nordic Optical Telescope, operated on the island of La Palma jointly by Denmark, Finland, Iceland,
 Norway, and Sweden, in the Spanish Observatorio del Roque de los Muchachos of the Instituto de
 Astrofisica de Canarias.}}
 
   \author{J.-E. Solheim  \inst{1} \and G. Vauclair \inst{2}  \and A. S. Mukadam\thanks{Hubble fellow} \inst{3}
  \and R. Janulis\inst{4} \and V. Dobrovolskas\inst{5}
          }
   \offprints{J.-E. Solheim}
\institute {Institute of Theoretical Astrophysics, University of Oslo, Box 1029 Blindern, N-0315 Oslo, Norway.\\
\email{j.e.solheim@astro.uio.no}
\and Universite Paul Sabatier, Observatoire Midi-Pyrenees, CNRS/UMR5572, 14 Av. Belin, 31400 Toulouse, France.\\
\email{gerardv@ast.obs-mip.fr}
\and Department of Astronomy, University of Washington, Seattle, WA 98195-1580, U.S.A.\\
\email{anjum@astro.washington.edu}
\and Institute of Theoretical Physics and Astronomy of Vilnius University, Gostauto 12, Vilnius LT-01108, Lithuania. \\
\email{jr@itpa.lt}
\and Observatory of  Vilnius University, M.K. Ciurlionio 29, Vilnius, LT-03100, Lithuania.\\
\email{vidas.dobrovolskas@ff.vu.lt}
}
   \date{Received 27 February 2007 /Accepted 11 April 2007 }
 
  \abstract
   {Most PG\,1159 stars are hydrogen deficient post-AGB stars on their way to the white dwarf 
cooling sequence. However, a fraction of them show small amounts of atmospheric hydrogen and are
 known as the hybrid PG\,1159 stars. Both the normal and hybrid PG\,1159 stars may reside at the
centre of a Planetary Nebula. Some PG\,1159 stars exhibit non-radial $g$-mode
pulsations, providing us with a means to constrain their fundamental physical parameters,
internal structure, and evolutionary status; asteroseismology is the technique of using global
stellar pulsations to probe the star's interior.}
 { Model calculations demonstrate that the PG\,1159 instability strip is driven by the partial
ionization of carbon and oxygen through the $\kappa$-mechanism. The range of excited frequencies in
 models is sensitive to the details of the chemical composition in the driving zone. Hydrogen content
plays a crucial role in deciding the range of excited frequencies for models of hybrid PG\,1159 pulsators.
The range of periods observed in all PG\,1159 pulsators can then be compared against the range predicted
 by theoretical models, which are based on observations of their chemical composition and evolutionary
 status. Our aim is to test the validity of the predicted range of pulsation periods for the hybrid PG\,1159
 variables, Abell\,43 and NGC\,7094.}
   {To achieve this goal, we acquired continuous high speed CCD photometry of the
hybrid PG\,1159 star Abell\,43 using the 2.6\,m Nordic Optical Telescope and the 3.5\,m
telescope at Apache Point Observatory. We also observed NGC\,7094, a hybrid PG\,1159 star with
atmospheric parameters similar to Abell\,43. We analyzed our data to obtain the pulsation spectra of the two
 stars.}
   {We detected six significant pulsations in the light curves of Abell\,43 with periods
between 2380\,s and 6075\,s. We were also fortunate to detect low amplitude pulsations in our
NGC\,7094 data with periods between 2000\,s and 5000\,s. The observed range of periods for
both stars is consistent with the theoretical predictions.}
   {The range of periods observed in both hybrid PG\,1159 stars, Abell\,43 and NGC\,7094, agree 
 with the theoretical model that the pulsations are driven by the
 $\kappa$-mechanism induced by the partial ionization of carbon and oxygen.}
   
\maketitle
%
%
\section{Introduction}
\begin{figure}
\resizebox{\hsize}{!}{\includegraphics{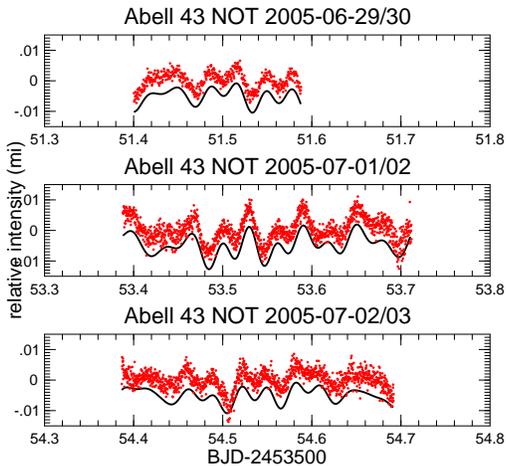}}
      \caption{Light curves of Abell\,43 shown for the three long NOT runs.
The solid lines are synthetic light curves constructed from the 
parameters listed in Table 2, and shifted downward by 0.005 units for clarity.}
     \label{Fig1}
\end{figure}

Hydrogen deficient hot white dwarf stars constitute the PG\,1159 class, and are often
found at the centre of planetary nebulae.
 \object {Abell\,43\,(=\,WD\,1751+10)} is one of the four known hybrid PG\,1159 stars, which
 show evidence of a non-negligible hydrogen abundance in their atmospheres (Napiwotzki \&
 Sch\"onberner \cite{napiwotzki}). Miksa et al. (\cite{miksa}) recently published the following atmospheric
 parameters for Abell\,43: ${\mathrm{T}}_{\mathrm{eff}}$\,=\,110\,kK,
log $g$\,=\,5.7, X(H)\,=\,0.35, X(He)\,=\,0.42, and X(C)\,=\,0.23 in mass fractions.
Werner \& Herwig (\cite{werner}) found an even higher carbon abundance from
HST ultra-violet observations of Abell\,43.

Low carbon abundance determinations by Dreizler et al. (\cite{dreizler}) had initially led
Quirion et al. (\cite{quirion04}) to rule out pulsations in Abell\,43 based on their model calculations.
But the recent high carbon abundance measurements led Quirion et al. (\cite{quirion05}) to
re-evaluate their previous conclusions about Abell\,43. They found that the revised models
suggested that Abell\,43 should exhibit $\ell$\,=\,1 $g$-mode pulsations, driven by the classical
$\kappa$-mechanism due to the ionization of carbon K-shell electrons, in the range of
 $\sim$2604--5529\,s. C\'orsico et al. (\cite{corsico}) carried out nonadiabatic pulsation calculations
for a grid of evolutionary PG\,1159 models. They predicted that Abell\,43 should exhibit periods in the range
 of $\sim$2600--7000\,s adopting a 0.530M$_{\sun}$ stellar mass evolutionary track.

Ciardullo \& Bond (\cite{ciardullo}) were the first to report possible variations in the
light curve of Abell\,43 with a period of about 2473\,s. Schuh et al. (\cite{schuh}) detected a
longer period of 5500\,s. They also suggested that the $\sim$2500\,s period reported
by Ciardullo \& Bond (\cite{ciardullo}) could simply be an artifact of the 5500\,s period caused by 
the short duration of the observations. Vauclair et al. (\cite{vauclair}) discovered
multiple modes in the pulsation spectrum of Abell\,43, and reported periods at 2600\,s and 3035\,s.
Their observations were based on three individual runs, amounting to a net observation time of 7.5\,hr
at the 2.6\,m Nordic Optical Telescope (NOT).

In this paper, we report the results from 24 hours of time-series photometry acquired on Abell\,43
in 2005 with the 2.6\,m NOT and the 3.5\,m telescope at Apache Point Observatory (APO).
We also report pulsation periods detected in the hybrid PG\,1159 star \object {NGC\,7094\,
 (=\,WD\,2134+125)}, which is almost like a spectroscopic twin of Abell\,43.
%
%
\section{Observations}
%
\begin{table}
\begin{minipage}[t]{\columnwidth}
\caption{Log of observations of Abell\,43}             
\label{tablelog}      
\centering  
\renewcommand{\footnoterule}{}  
\begin{tabular}{c c c c r c }     
\hline\hline                   
Date & Start & Length& Cycle&Data& Observer\\
         & (UT)  &(hrs)& time(s)&points&\\ 
\hline   
        2005-06-29  & 21:28     &4.40 & 30 &  541 & JES\footnote{J.~-~E.~Solheim}
  \\
  2005-07-01 & 03:43  &1.56  & 20 &  283 & JES
 \\
  2005-07-01  & 09:14     &1.54& 37 &  51 & ASM\footnote{A.~S.~Mukadam}
 \\
  2005-07-01  & 21:08      &7.71& 20 & 1399 & JES
  \\
  2005-07-02  & 21:07       &7.32& 20&1319 & JES
   \\
  2005-07-10  & 07:24      & 1.71&41 & 110& ASM
  \\     
\hline    
\end{tabular}
\end{minipage}
\end{table}
%
%
We observed Abell\,43 for four nights, from the 29th of June to the 2nd of July, 2005 using the 2.6\,m NOT
in addition to two short runs with the 3.5\,m telescope at APO on the 1st and 10th of July, 2005.
The log of observations is given in Table~\ref{tablelog}.

We conducted observations at the NOT using the Andalucia Faint Object Spectrograph and Camera
 (ALFOSC), which is equipped with a thinned 2048\,$\times$\,2048 E2V CCD\,42-40 chip.
We observed with the broadband blue W92 filter, which has a FWHM of 275\,nm centered at 550\,nm; this
enables us to gather a fair amount of flux from the target and minimize the sky contribution at the same
 time. We acquired suitable flat fields each night just after sunset and determined the dark current
for each individual image using the overscan region. We controlled the ALFOSC data acquisition using the
 software interface \emph{tcpcom}, defining multiple windows to achieve fast readouts (\O stensen
 \cite{ostensen}). We achieved a net readout time of 4.8 to 5.8\,s for the six
separate windows around the target star, three comparison stars, and two blank fields for determining the
 simultaneous sky brightness. We chose a time resolution of 30\,s during our first night at the NOT
 telescope, and our exposure time became 23.9\,s. We reduced the cycle time to 20\,s on subsequent
 nights, and achieved exposure times of about 14.1, 14.8, and 14.9\,s for slightly different windowing
 patterns. With real time processing, we were able to display light curves of the raw and sky-subtracted data
 during acquisition.
 
We followed the standard basic reduction procedure, subtracting a dark frame from each image and then
 flat-fielding it. After these preliminary steps, we computed light curves using aperture photometry for
 several aperture sizes. We selected the best light curve based on the highest S/N ratio. For the
 observations outlined above, we obtained the best result for an aperture diameter of 24 pixels
corresponding to 4.6\,arcseconds on the sky.
We corrected for transparency variations and differential extinction by dividing the target star
light curve with the average light curve of the comparison stars.
In Fig. \ref{Fig1} we show the light curves from the three longest NOT runs.

The APO observations were acquired with the Seaver Prototype Imaging camera (SPIcam) and a Johnson B
 filter. We used 3x3 binning and windowing in order to reduce the read out time to about 25--35\,s. The
 instrument has a read noise of 5 electrons RMS, a gain of 1 e/ADU, and a plate scale of 0.42 arcsec/pixel
 with 3x3 binning. We used a standard IRAF reduction to extract sky-subtracted light curves from the CCD
 frames using weighted circular aperture photometry (O'Donoghue et al. \cite{odonoghue}).

\section{Hybrid PG\,1159 star Abell\,43.}
 \begin{figure}
\resizebox{\hsize}{!}{\includegraphics{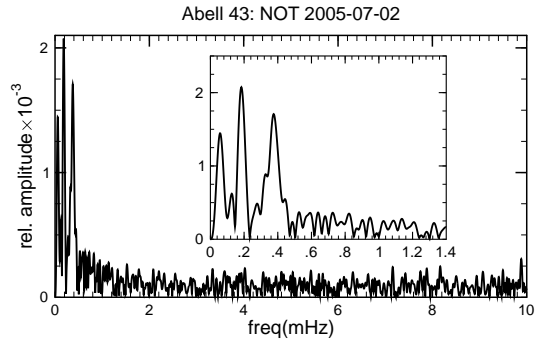}}
    \caption{Pulsation spectrum of the light curve of the NOT run
from the 2nd of July, 2005. The insert is a blow up of the low
 frequency region, that shows the evident lack of excited frequencies 
above 0.5\,mHz.}
    \label{Fig2}
 \end{figure}
\begin{table*}
\begin{minipage}[t]{\columnwidth}
\caption{Identified periodic signals in Abell\,43 in the three long NOT runs}             
\label{tableruns}      
\centering  
\renewcommand{\footnoterule}{}  
\begin{tabular}{c c r r c r r c r r c r r c c}     
\hline\hline       
Date & Res.&F1 & P1&A1 &F2&P2&A2&F3&P3&A3&F4&P4&A4&F3/F1\\
2005&($\mu$Hz)&($\mu$Hz)&(s)&(mma\footnote{relative amplitude$\times$10$^{-3}$})&($\mu$Hz)&(s)&(mma)&($\mu$Hz)&(s)&(mma)&($\mu$Hz)&(s)&(mma)\\
\hline   
06-29&63&183.3&5456&2.0&&&&363.9&2748&2.3&416.3&2402&0.8&1.98 \\
$\sigma$&&1.4&41&0.1&&&&1.2&10&0.1&3.6&21&0.1\\
07-01&36&175.5&5698&3.1&321&3118&1.0&366.4&2729&2.7&408.3&2449&1.4&2.08\\
$\sigma$&&1.2&38&0.1&3.4&34&0.1&0.9&10&0.1&3.4&15&0.1\\
07-02&38&185.2&5399&2.1&325&3075&0.8&368.8&2711&1.9&418.9&2387&2.0&1.99\\
$\sigma$&&1.6&47&0.1&4.4&41&0.1&1.8&1.4&0.1&3.3&19&0.1\\   
\hline    
\end{tabular}
\end{minipage}
\end{table*}
%
%
\subsection {Single night observations}
We initially computed Discrete Fourier Transforms (DFTs) of the three long light curves acquired using the
 NOT. We show only the DFT from one of the longest runs in Fig. \ref{Fig2}, as the other DFTs look similar.
We list the following observed features evident from the DFTs of our NOT data:
\begin{itemize}
\item
Each pulsation spectrum shows three dominant components, which are relatively stable in frequency,
but exhibit  variable amplitudes.
\item All pulsation spectra have a strong high frequency cut-off above 0.5 mHz, beyond which we do not
 detect any signals.
\item The light curves are non-stationary\footnote{When signals detected in a light curve are not present for
its entire duration, it is said to be non-stationary in nature.} and we have to be cautious
in drawing conclusions using standard tools for spectral analysis.
\end{itemize}

\begin{figure}
\resizebox{\hsize}{!}{\includegraphics{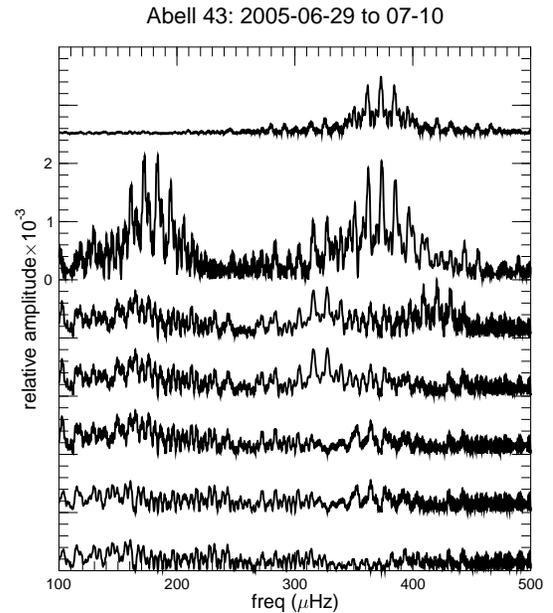}}
   \caption{We show the DFT computed from the combined light curve of five runs
used in the analysis. The top curve shows the window
 function. The second curve shows the DFT of the light curve. The third curve shows the DFT after
 subtracting out the two highest amplitude peaks. The lower four curves show the DFT after subtracting
 three, four, five, and six frequencies respectively.}
     \label{finalFT}%
\end{figure}

The dramatic cut off implies that we only find pulsations with periods longer than $ \sim$2000\,s, as
 predicted both by Quirion et al. (\cite{quirion05}) and C\'orsico et al. (\cite{corsico}).
We determined the frequencies, amplitudes, and phases of the significant peaks in the pulsation spectra
using a simultaneous non-linear least squares fit (\emph{Period04}; Lenz \& Breger \cite{lenz}).
To determine which peaks in the DFT constituted real power, we used the method
of prewhitening in conjunction with the non-linear least squares analysis.
This involves fitting the dominant component and subtracting
it from the light curve. If we now compute a DFT of this resultant light curve,
the peak corresponding to the dominant mode will be absent. We repeat
the procedure for the next highest peak in the pre-whitened DFT, fitting both
high-amplitude components simultaneously and then subtracting them
from the light curve. The resultant DFT will not show
power due to both the highest and second-highest peaks.
We continue this process as long as there are well-resolved
 peaks in the DFT. When the prewhitened DFT does not
show any significant power, then we can be sure that we have determined the
entire pulsation spectrum using the simultaneous non-linear least squares fit.
Fig. \ref{finalFT} serves as a demonstration of the prewhitening method.
 
We show the results of our analysis for the three longest NOT runs
in Table \ref{tableruns}, that list the significant periods detected with a False Alarm Probability better
than 1/100 (Scargle\cite{scargle}). The uncertainties listed in the table are 1\,$ \sigma $ errors from the
non-linear least squares fit. Fig. \ref{Fig1} also includes the synthetic light curves constructed from the
frequencies, amplitudes and phases determined as above.

It is evident from Table \ref{tableruns} that F1 varies most in frequency and amplitude from night to night.
 These variations may be caused by unresolved power.
We also notice that the ratio of the frequencies F3:F1 is very
close to two on the first and third nights, insinuating a harmonic ratio. The ratio is close to 2.08 on the
second night and may perhaps be due to F1 being an unresolved multiplet.

%
%
%
If we compare the periods we detect to those observed earlier, we find that the period 2473\,s discovered
by Ciardullo \& Bond(\cite{ciardullo}) is close to P4, while the period of 5500\,s found by Schuh et al.
 (\cite{schuh}) is close to P1. The two periods 2600\, and 3035\,s, discovered by Vauclair et al.
 (\cite{vauclair})  are not too far from P3 and P2. The discrepancies may either be due to insufficient
resolution or may perhaps indicate real period changes between the different pulsation modes.
\subsection{Combining all useful observations}
To investigate whether the nightly
variations in frequencies and amplitudes were due to unresolved peaks,
we re-analyzed the combined light curves with the addition of the 
APO observations from the 10th of July, 2005.
Note that we could not utilize the APO observations from the 1st of July as bad
weather rendered them far too
noisy for inclusion in our analysis.

We computed a combined DFT again, and used the prewhitening technique
in conjunction with the non-linear least squares analysis to
arrive at the values listed in Table \ref{tablecomb}.
We show our final DFT in Fig. \ref{finalFT}, along with the different stages
of prewhitening. We find that the power seen earlier as  F1 split into two individual frequencies f1 and f2,
 while the power at  F3 split into f4 and f5. Using the new frequency components, we find that {
 f3/f1\,=\,1.991 and (f4\,+\,f5)/2\,f2\,=\,2.005, giving us two ratios of frequencies close to 2.
%
\begin{table}
\begin{minipage}[t]{\columnwidth}
\caption{Identified periodic signals in Abell\,43 from the combined data set}             
\label{tablecomb}      
\centering  
\renewcommand{\footnoterule}{}  
\begin{tabular}{c r r c c }     
\hline\hline                    
Name&  Freq & Period & Amplitude &Noise\\
&  (${\mu}$Hz) & (s) & (mma\footnote{relative amplitude$\times$10$^{-3}$})&(mma) \\
\hline 
f1&164.62& 6075 & 0.8 & 0.15\\  
$\sigma$&0.08&3&0.1\\
f2 & 183.76& 5442 & 1.9&0.15 \\
$\sigma$&0.03&1.0&0.1\\
f3& 327.70  &3051.6 &0.9&0.20 \\
$\sigma$&0.07&0.7&0.1\\
f4& 363.79& 2748.8 & 0.9&0.20\\
$\sigma$&0.08&0.6&0.1\\
f5&373.12&2680.1&2.0&0.20\\
$\sigma$&0.03&0.2&0.1\\
f6&420.23&2379.7&1.0&0.20\\
$\sigma$&0.06&0.4&0.1\\
\hline    
\end{tabular}
\end{minipage}
\end{table}

Averaging the observations to a sampling time of 200\,s, we show the observed
light curves along with the synthetic light curves constructed from our fit
in Fig. \ref{lcp1}; an inspection of this figure reveals that the fit is
quite good. This tells us that Abell\,43 is more stable than the pulsating PNNi with shorter
 periods, investigated by Gonz\'alez P\'erez et al. (\cite{perez}). We also notice how the light curve can
 change dramatically on short time scales due to beating of the closely spaced modes; this suggests
 observations over a shorter time span compared to the beat cycle may give wrong results or show no
 pulsations at all.

 \begin{figure}
\resizebox{\hsize}{!}{\includegraphics{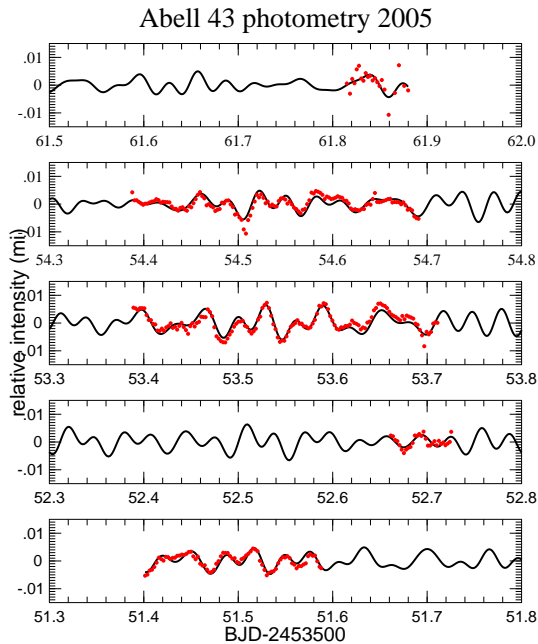}}
   \caption{We show observed light curves, averaged over a sampling time of 200\,s,
compared with synthetic curves  constructed from parameters in Table \ref{tablecomb}.}
  \label{lcp1}%
\end{figure}
%
%
The shortest period we detect in our data is 2380\,s, which is quite close to the predicted lower limit of
 2600\,s for $\ell$\,=\,1 $g$-mode pulsations induced by the
 $\kappa$-mechanism  (Quirion et al. \cite{quirion05} and C\'orsico et al. \cite{corsico}). The longest period
 of 6075\,s is slightly longer than predicted by Quirion et al. (\cite{quirion05}) but within the range predicted
 by C\'orsico et al. (\cite{corsico}). The non stationary nature of the light curves makes it difficult to draw
 certain conclusions about the frequency ratios close to 2.00. They may indicate simple harmonics or non
 linear resonances as proposed by Buchler et al. (\cite{buchler}).} 
\section{Hybrid PG\,1159 star NGC\,7094: Almost a spectroscopic twin of Abell\,43}
\begin{figure}
\resizebox{\hsize}{!}{\includegraphics{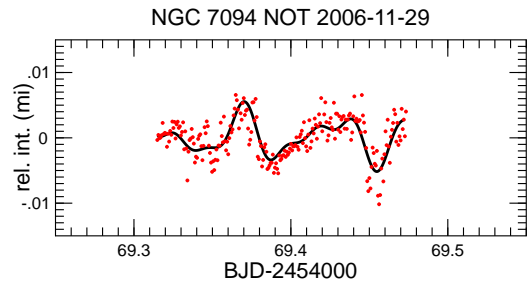}}
   \caption{We show the observed light curve of NGC\,7094 with a simulated light curve based on the
 periods 34, 50, and 83\,min.}
      \label{n7094fig}
\end{figure}
The hybrid PG\,1159 star NGC\,7094 shows almost the same spectroscopic parameters as Abell\,43,
except that 1\% He in Abell\,43 has to be replaced by 1\% O (Miksa et al. \cite{miksa}). Quirion et al.
 (\cite{quirion05}) propounded that NGC\,7094 should pulsate with periods in the range of 2550\,-\,5413\,s.
Previous searches by Ciardullo et al. (\cite {ciardullo}) and
 Gonz\'alez P\'erez et al. (\cite{perez}) did not reveal any pulsations in the star.
The latter observation lasted only 5220\,s and
 was probably too short to detect pulsations in the predicted range. 

In order to scrutinize the variability of NGC\,7094, we asked for a service run on the NOT, which
can be a maximum of 4\,hr long. We obtained a light curve of length 13692\,s with ALFOSC on the 29th of
 November, 2006. We used the B filter and a window of size 1k\,$ \times $\,1k pixels. We
chose an exposure time of 30\,s and the readout time was about the same,
giving us a time resolution of 60\,s. Fig. \ref{n7094fig} shows the light curve we obtained from our
 observations. With the false alarm probability criterion of 1/100 or better, we identified three peaks with
 periods between 2000\,s and 5000\,s. A synthetic light curve constructed from these peaks is also shown
 in Fig. \ref{n7094fig}. Longer observing runs are needed to  characterize the pulsations better, but this
 paper serves to establish NGC\,7094 as a low-amplitude pulsator.

The range of pulsation periods is shifted toward  shorter periods relative to Abell\,43. This is predicted by
 Quirion et al. (\cite{quirion05}) and is a consequence of adding 
a trace of oxygen in their models. The observed lower period limit of 2000\,s
is shorter than the theoretical limit of 2550\,s predicted by the same authors.
\section{Conclusions}
For the hybrid PG\,1159 star Abell\,43, we find six pulsation periods with a range close to that predicted for
 $\ell$\,=\,1 $g$-mode pulsations driven by the $\kappa$-mechanism due to partially
 ionized carbon (Quirion et al. \cite{quirion05}; C\'orsico et al. \cite{corsico}). The light curves are non
 stationary, but the pulsation frequencies and amplitudes are relatively stable during the 11 day span of the
 observations. This is different from pulsating PNNi, which showed changes in amplitudes from night to
 night, and also  during the same night (Gonz\'alez Per\'ez et al. \cite{perez}).

We have also detected pulsations in the hybrid PG\,1159 star NGC\,7094 which is an approximate
 spectroscopic twin of Abell\,43. The observed pulsations lie in the range of 2000--5000\,s, in a band of
 periods shifted to shorter values than for Abell\,43. This agrees with theory and occurs when a small
 amount of oxygen is added to the model pulsator (Quirion et al. \cite{quirion05}), although the lowest period
 observed is somewhat shorter than predicted for NGC\,7094. 

\begin{acknowledgements}
The data presented here have mostly been acquired using ALFOSC, which is owned by the Instituto de
 Astrofisica de Andalucia (IAA) and operated at the Nordic Optical Telescope under agreement between IAA
 and the NBIfAFG of the Astronomical Observatory of Copenhagen. We would like to specially thank
the NOT staff for their service observations of NGC\,7094 resulting in the detection of a new
 pulsator. The rest of the data are based on observations obtained with the Apache Point Observatory
 3.5-meter telescope, which is owned and operated by the Astrophysical Research Consortium. 
Support for a fraction of this work was provided by NASA through the Hubble Fellowship grant
HST-HF-01175.01-A awarded by the Space Telescope Science Institute, which is operated
by the Association of Universities for Research in Astronomy, Inc., for NASA, under contract NAS 5-26555.
We also thank the anonymous referee for suggesting a number of improvements to the first version of this
 manuscript.

\end{acknowledgements}

\end{document}